\documentclass{aa}
\usepackage{graphics,epsfig,txfonts}
\usepackage{color}
\usepackage{natbib}
\usepackage[colorlinks=true, citecolor=blue]{hyperref}
\bibpunct{(}{)}{;}{a}{}{,}

\newcommand{\ergcm}[1]{$\times 10^{#1}$ erg cm$^{-2}$ s$^{-1}$}

\newcommand{\ergs}[1]{$\times 10^{#1}$ erg s$^{-1}$}
\newcommand{\oergs}[1]{$10^{#1}$ erg s$^{-1}$}
\newcommand{\hcm}[1]{$\times 10^{#1}$ cm$^{-2}$}
\newcommand{\ohcm}[1]{$10^{#1}$ cm$^{-2}$}
\newcommand{\expo}[1]{$\times 10^{#1}$}
\newcommand{\oexpo}[1]{$10^{#1}$}

\newcommand{\nh}{N$_{\rm H}$\xspace}

\newcommand{\lbol}{\hbox{L$_{\rm bol}$}}

\newcommand{\ct}{{cts s$^{-1}$}\xspace}

\newcommand{\Hone}{\ion{H}{i}\xspace}

\newcommand{\ltsima}{$\buildrel < \over \sim$}
\newcommand{\lsim}{\lower.5ex\hbox{\ltsima}}
\newcommand{\gtsima}{$\buildrel > \over \sim$}
\newcommand{\gsim}{\lower.5ex\hbox{\gtsima}}
\newcommand{\xmm}{XMM-{\it Newton}\xspace}

\newcommand{\ein}{{\it Einstein}\xspace}
\newcommand{\srg}{{\it SRG}\xspace}
\newcommand{\ero}{\mbox{eROSITA}\xspace}
\newcommand{\swift}{{\it Swift}\xspace}
\newcommand{\rosat}{{\it ROSAT}\xspace}
\newcommand{\gaia}{{\it Gaia}\xspace}
\newcommand{\xspec}{\texttt{XSPEC}\xspace}
\newcommand{\eSASS}{\texttt{eSASS}\xspace}
\newcommand{\smplmcn}{\mbox{SMP\,LMC\,21}\xspace}
\newcommand{\sssa}{\mbox{XMMU\,J050452.0$-$683909}\xspace}
\newcommand{\sssb}{\mbox{XMMU\,J051854.8$-$695601}\xspace}
\newcommand{\sssc}{\mbox{XMMU\,J050815.1$-$691832}\xspace}
\newcommand{\sssd}{\mbox{XMMU\,J044626.6$-$692011}\xspace}
\newcommand{\ssca}{\mbox{050452.0$-$683909}}
\newcommand{\sscb}{\mbox{051854.8$-$695601}}
\newcommand{\sscc}{\mbox{050815.1$-$691832}}
\newcommand{\sscd}{\mbox{044626.6$-$692011}}
\newcommand{\ssa}{\mbox{J0504}\xspace}
\newcommand{\ssb}{\mbox{J0518}\xspace}
\newcommand{\ssc}{\mbox{J0508}\xspace}
\newcommand{\ssd}{\mbox{J0446}\xspace}
\newcommand{\smpsmc}{\mbox{SMP\,SMC\,22}\xspace}
\newcommand{\xmmdd}{\mbox{3XMM\,J051034.6$-$670353}\xspace}
\newcommand{\xmmlmco}{\hbox{XMMU\,J050803.1$-$684017}\xspace}
\newcommand{\smplmco}{\hbox{SMP\,LMC\,29}\xspace}

\begin{document}
 
\title{Discovery of four super-soft X-ray sources in XMM-Newton observations of the Large Magellanic Cloud}

\author{C. Maitra\inst{\ref{mpe}} \and
        F. Haberl\inst{\ref{mpe}}
       }

\titlerunning{Super-soft X-ray sources in XMM-Newton observations of the LMC}
\authorrunning{Maitra et al.}

\institute{Max-Planck-Institut f\"ur extraterrestrische Physik, Gie{\ss}enbachstra{\ss}e 1, 85748 Garching, Germany\label{mpe}, \email{fwh@mpe.mpg.de}
          }

\date{Received DD Month 2021 / Accepted DD Month 20..}

\abstract
  {Super-soft X-ray sources were established as a heterogeneous class of objects from observations of the Large Magellanic Cloud (LMC).}
  {We have searched for new sources of this class in the X-ray images obtained from the \xmm survey of the LMC and additional archival observations.}
  {We first selected candidates by visual inspection of the image, and screened out artefacts which can mimic super-soft X-ray sources as well as bright foreground stars which create optical loading on the CCD image. We finally obtained 4 new super-soft X-ray sources for which we performed detailed X-ray timing and spectral analysis and searched for possible optical counterparts to identify their nature. We also looked at archival \rosat and \swift observations to investigate the long-term behaviour of the sources.}
  {\sssa is identified as the central star of the planetary nebula \smplmcn in the LMC. We suggest \sssb and \sssc as new soft intermediate polars from the nature of their X-ray spectrum. Their estimated absorption-corrected luminosities and the blackbody radii indicate that they are located in our Galaxy, rather than the LMC. We discovered coherent pulsations of 497\, s from \sssd which indicates a magnetic cataclysmic variable nature of the source. The location of \sssd in the LMC or our Galaxy is less clear. It could either be a white dwarf in the LMC with nuclear burning on its surface near the Eddington limit, or another soft intermediate polar in our Galaxy.}
  {The discovery of new super-soft X-ray sources makes a significant contribution to the known population in our own Galaxy. 
  An observed higher density of sources in the direction of the Magellanic Clouds can likely be explained by the relatively low Galactic column density in their direction as well as a large number of existing observations sensitive at low X-ray energies.}

 
\maketitle
 
\section{Introduction}
\label{sec:introduction}

The low Galactic foreground absorption in the direction of the Magellanic Clouds makes them ideal laboratories
for the detection and investigation of super-soft X-ray sources (SSSs). In fact, the first SSSs were discovered 
in the Large Magellanic Cloud (LMC) using the \ein observatory \citep{1981ApJ...248..925L} before further discoveries with \rosat 
established them as a new, though heterogeneous, class of objects 
\citep[see][]{1996LNP...472..299G,2006A&A...452..431K,2008A&A...482..237K}.

SSSs are characterised by soft X-ray spectra with kT $\sim$ 15$-$80 eV \citep[e.g.][]{1997ARA&A..35...69K} and a 
wide range of luminosities. The most luminous ($\sim$10$^{36}$ to $\sim$\oergs{38}) can be explained by stable nuclear 
burning white dwarfs (WDs), which accrete H-rich matter from a companion star \citep{1992A&A...262...97V}. 
The size of the emission area derived from blackbody fits to their spectra is consistent with emission from the 
full WD surface.
At least 5 of these close-binary SSSs are known in the LMC \citep{2008A&A...482..237K}. 
WDs as central stars of planetary nebulae are known as SSSs in the LMC \citep[\smplmco,][]{2008A&A...482..237K} and 
the Small Magellanic Cloud \citep[SMC; \smpsmc and \hbox{SMP\,SMC\,25};][]{2010A&A...519A..42M}.

A soft blackbody-like emission component is also observed from magnetic cataclysmic variables (mCVs), however, with 
significantly lower luminosities below \oergs{34}. 
AM\,Her-type systems \citep[polars,][]{1990SSRv...54..195C} and DQ\,Her-type systems
\citep[soft intermediate polars,][]{1995A&A...297L..37H,1996A&A...310L..25B,2008A&A...489.1243A} show pulsations in their X-ray flux, caused  
by the rotation of the magnetised WD, which is synchronous with the orbital revolution in polars (typically longer 
than one hour). Typical WD spin periods of a few hundred seconds \citep{2006csxs.book..421K} are found from intermediate polars (IPs) with P$_{\mathrm{spin}} \sim 0.1$P$_{\mathrm{orb}}$ \citep{2004ApJ...614..349N}.
Double-degenerate binaries, i.e. systems hosting two interacting WDs, can have orbital periods as short as a few minutes
\citep{1998MNRAS.293L..57C,2014A&A...561A.117E}.

In this paper we report the discovery of new SSSs in \xmm observations of the LMC.
In Sect.~\ref{sec:xray} we describe the \xmm and other available archival X-ray observations and the analysis methods. Sect.~\ref{sec:ogle} examines the OGLE data which cover the X-ray positions.
The results from our temporal and spectral analyses, and search for optical counterparts are presented in Sect.~\ref{sec:results} and discussed in Sect.~\ref{sec:discussion}.

\section{X-ray observations and analysis}
\label{sec:xray}

\subsection{\xmm}
\label{sec:XMM}

In the course of the \xmm surveys of the SMC \citep{2012A&A...545A.128H,2013A&A...558A...3S} 
and the LMC \citep{2016A&A...585A.162M} we searched for new SSS in the data obtained by the 
European Photon Imaging Cameras (EPIC), including pn- \citep{2001A&A...365L..18S} and 
MOS-type \citep{2001A&A...365L..27T} CCD detectors. Due to their soft X-ray spectra with the 
bulk of the emission below $\sim$1\,keV they appear as red sources in our RGB images, which are 
composed of combined EPIC images in three energy bands (R: 0.2--1.0\,keV, G: 1.0--2.0\,keV and 
B: 2.0--4.5\,keV). The images were produced using the same pipeline as for the SMC as described in \citet{2013A&A...558A...3S}.

However, CCD-detectors can also produce artefacts which mimic SSSs: 
1) hot CCD pixels or columns create events in low detector channels and 
2) optically bright stars produce optical loading due to the sensitivity of the CCDs to 
optical photons which can release a sufficiently high number of electrons to create a signal 
above the low-energy threshold of the detector.
Both of these types of false detections can be recognised, type 1) because they are not 
subject of blurring due to the point spread function (PSF) of the telescope and can be identified 
in detector images, type 2) correlate with optically bright stars and produce large numbers of 
invalid event patterns.

\begin{table*}
\centering
\caption[]{\xmm observations of new SSSs.}
\begin{tabular}{ccccccccc}
\hline\hline\noalign{\smallskip}
\multicolumn{2}{c}{Source name} &
\multicolumn{1}{c}{Observation} &
\multicolumn{1}{c}{Start time} &
\multicolumn{1}{c}{Exp.} &
\multicolumn{1}{c}{Off-axis} &
\multicolumn{1}{c}{R.A.} &
\multicolumn{1}{c}{Dec.} &
\multicolumn{1}{c}{Err} \\
\multicolumn{1}{c}{XMMU\,J...} &
\multicolumn{1}{c}{short} &
\multicolumn{1}{c}{ID} &
\multicolumn{1}{c}{} &
\multicolumn{1}{c}{} &
\multicolumn{1}{c}{angle} &
\multicolumn{2}{c}{(J2000)} &
\multicolumn{1}{c}{} \\
\multicolumn{2}{c}{} &
\multicolumn{1}{c}{} &
\multicolumn{1}{c}{} &
\multicolumn{1}{c}{(ks)} &
\multicolumn{1}{c}{(\arcmin)} &
\multicolumn{1}{c}{(h m s)} &
\multicolumn{1}{c}{(\degr\ \arcmin\ \arcsec)} &
\multicolumn{1}{c}{(\arcsec)} \\
\noalign{\smallskip}\hline\noalign{\smallskip}
\ssca & \ssa & 0803460101 & 2017-10-19 04:34 & 44.41 &  7.7  & 05 04 52.00 & -68 39 09.7 &  0.41 \\
      &      & 0693450201 & 2013-02-09 13:14 & 11.40 & 13.9  & 05 04 51.39 & -68 39 09.9 &  0.64 \\
\sscb & \ssb & 0690751801 & 2012-12-28 21:30 & 22.95 &  5.5  & 05 18 54.82 & -69 56 01.5 &  0.40 \\
\sscc & \ssc & 0690752001 & 2012-09-22 02:29 & 22.05 & 13.8  & 05 08 15.14 & -69 18 32.4 &  0.43 \\
\sscd & \ssd & 0801990301 & 2017-09-28 21:50 & 21.54 &  5.1  & 04 46 26.62 & -69 20 11.9 &  0.48 \\
\noalign{\smallskip}\hline
\end{tabular}
\tablefoot{
Parameters are given for the EPIC-pn instrument. Source coordinates were determined by applying bore-sight correction using background AGN \citep[following][]{2019A&A...622A..29M}.
The position error includes statistical and remaining systematic uncertainties (0.33\arcsec).
Exposure times are given after correcting for dead times and removal of intervals with high background.
}
\label{tab:observations}
\end{table*}

\begin{table}
\centering
\caption[]{\ero observations of \sssb}
\begin{tabular}{lrrrrrrr}
\hline\hline\noalign{\smallskip}
\multicolumn{1}{c}{Observation} &
\multicolumn{1}{c}{Start time} &
\multicolumn{1}{c}{Exp. (ks)} \\
\noalign{\smallskip}\hline\noalign{\smallskip}

 700182 & 2019-11-27 06:07:32.9 & 31.5  \\
 700201 & 2019-12-02 12:45:50.4 & 11 \\
 700202 & 2019-12-04 02:46:43.5  & 9 \\
\noalign{\smallskip}\hline
\end{tabular}
\label{tab:observations-ero}
\end{table}

\begin{figure}
 \resizebox{0.495\hsize}{!}{\includegraphics[clip=]{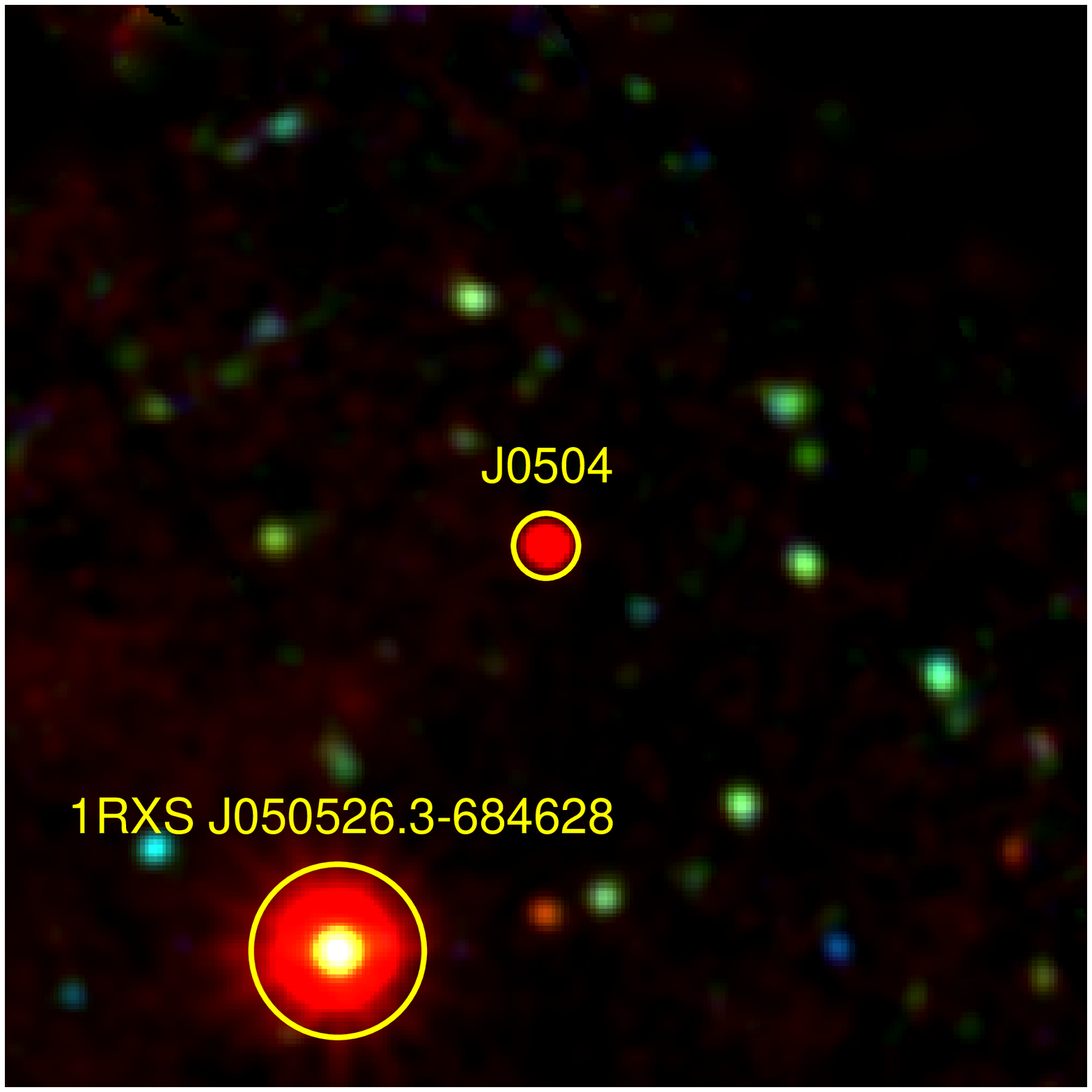}}
 \resizebox{0.495\hsize}{!}{\includegraphics[clip=]{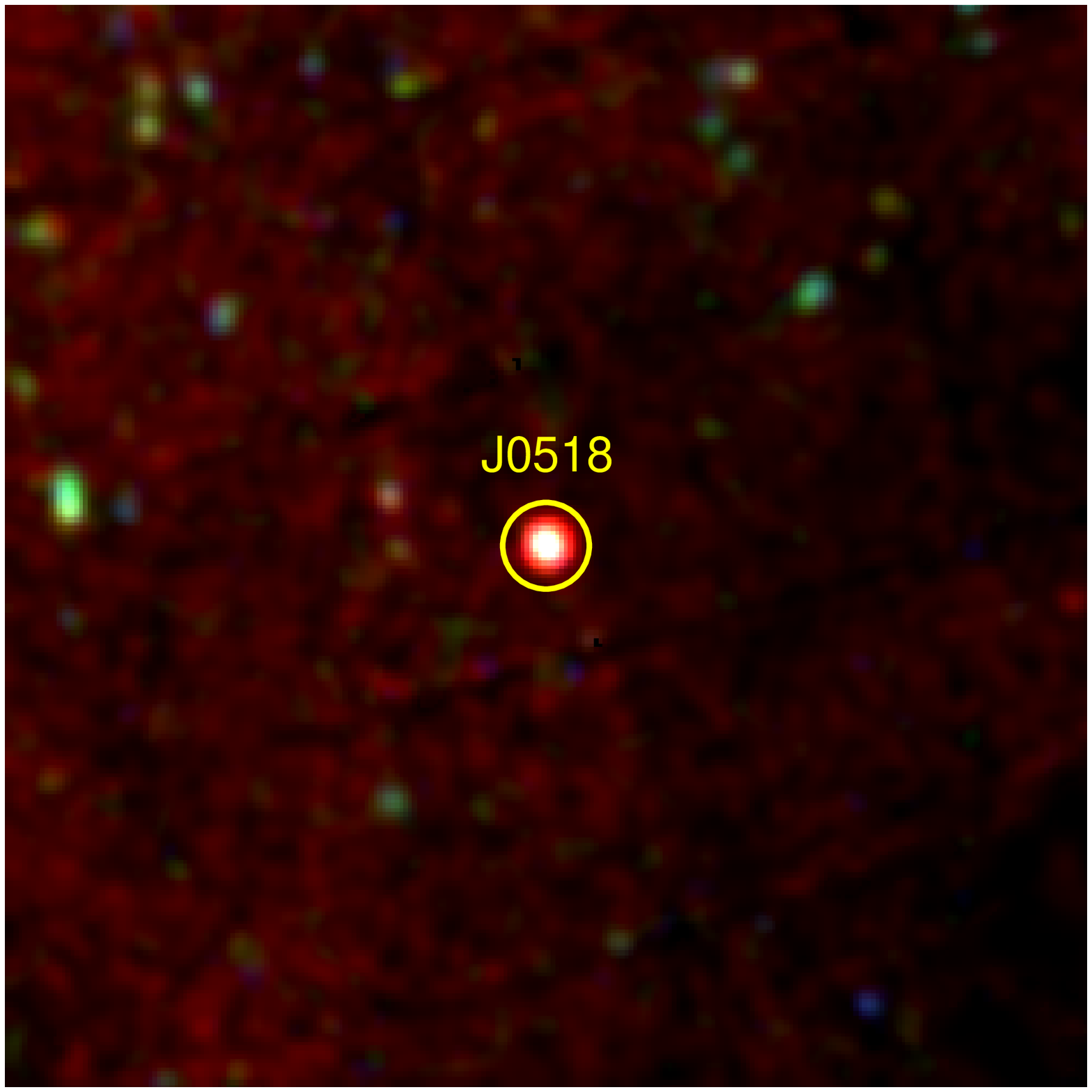}}\\
 \resizebox{0.495\hsize}{!}{\includegraphics[clip=]{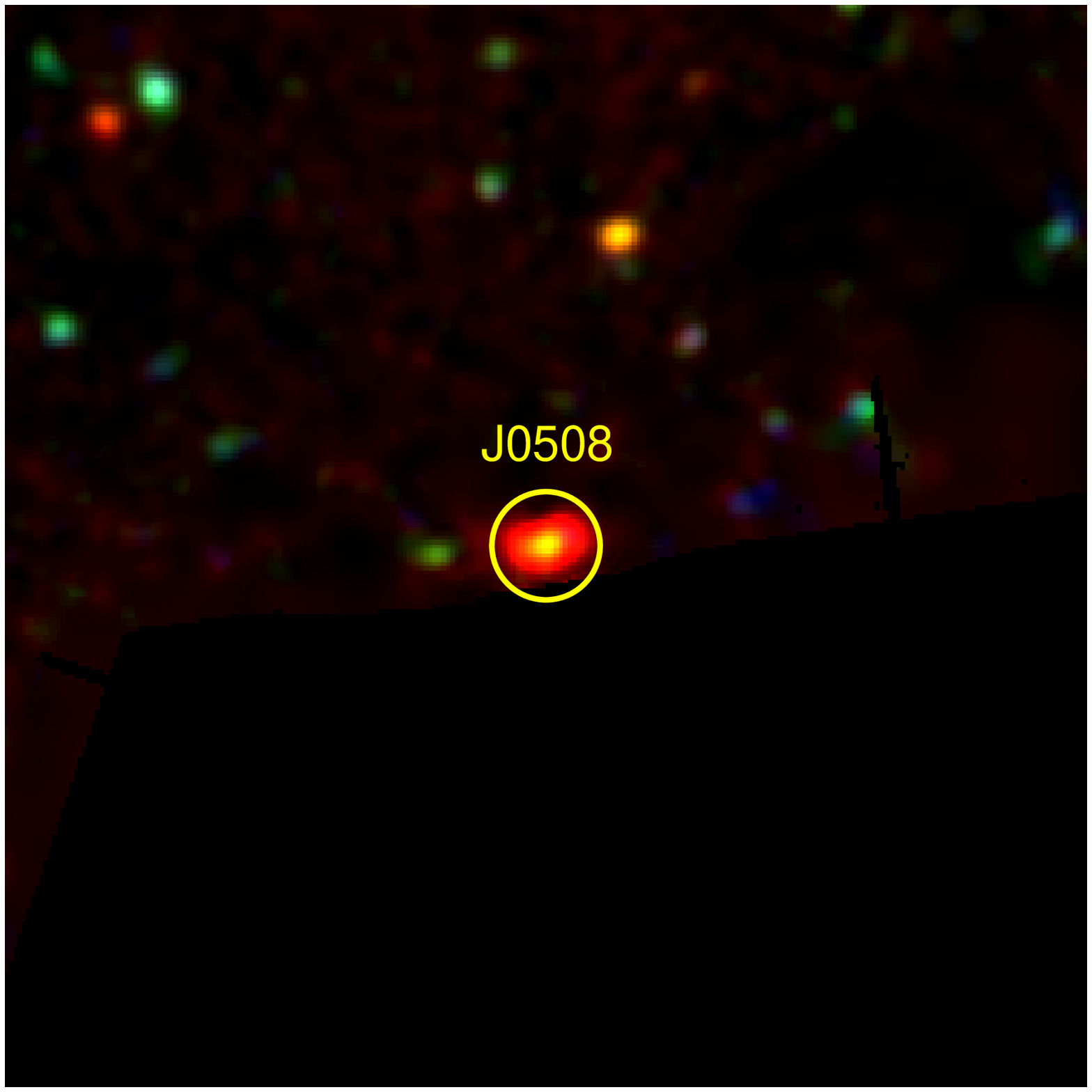}}
 \resizebox{0.495\hsize}{!}{\includegraphics[clip=]{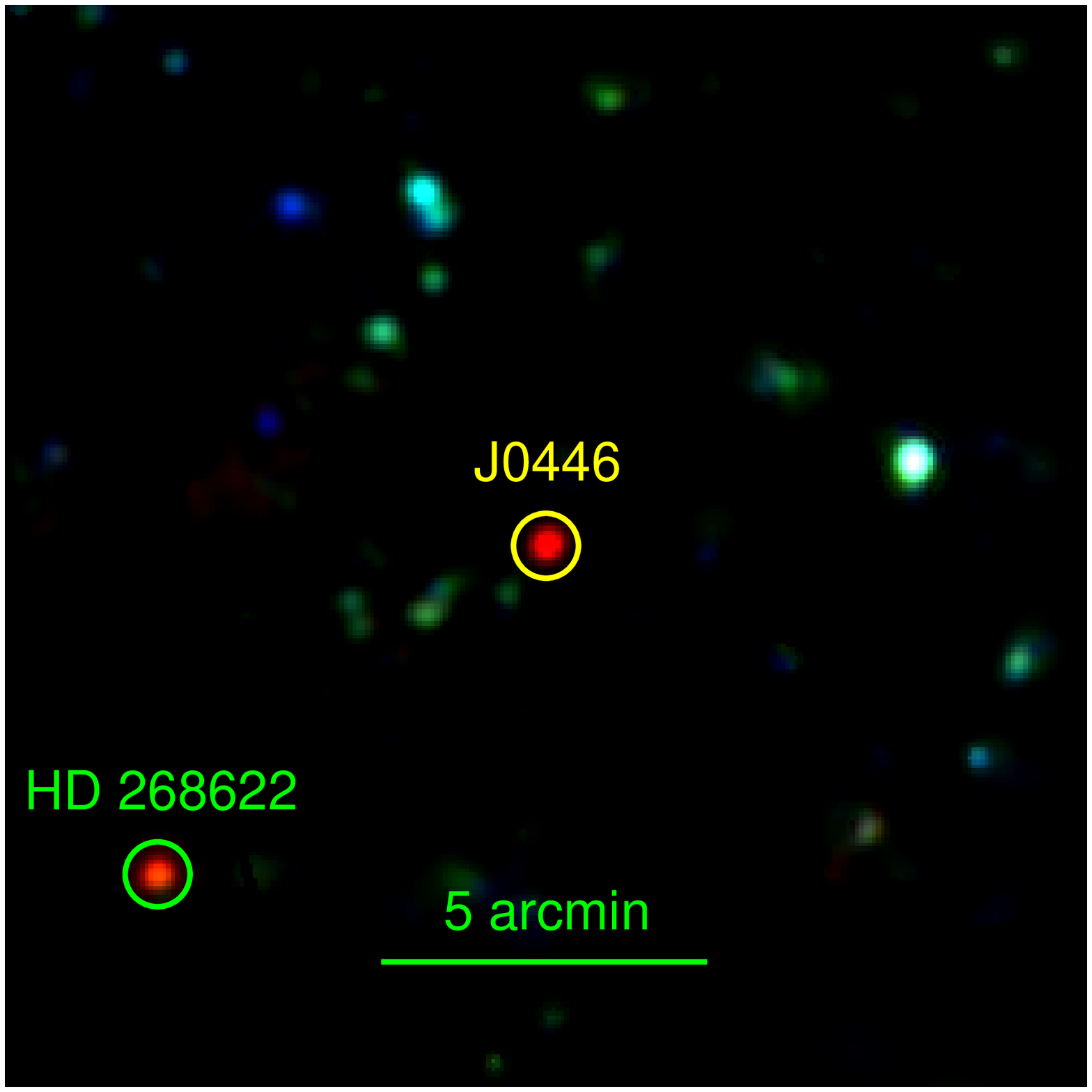}}
 \caption{
  The regions around the 4 new SSSs. The zoomed-in images are all on the same scale and extracted from the full RGB mosaic of the EPIC LMC observations. HD\,268622 is a bright foreground star of spectral type G5 and V = 11.3 mag, causing optical loading (see Sect.\,\ref{sec:XMM}).
 }
 \label{fig:SSSima}
\end{figure}

We processed all EPIC observations in the directions of the Magellanic Clouds using the \xmm 
SAS\,19.0.0\footnote{Science Analysis Software (SAS), \url{https://www.cosmos.esa.int/web/xmm-newton/sas}}
software package to create event lists using the most recent software and calibration files. Following 
\citet{2013A&A...558A...3S} we produced the images and run source detection.
From a visual inspection of the RGB images of the LMC we identified 22 candidates for new SSSs. 
In the majority of cases we identified a bright foreground star (V magnitudes were between 4.8 and 11.4) 
in observations for which the thin optical blocking filter was used, which was insufficient to suppress the 
optical light. Only in one case we found a hot pixel which was not masked out automatically by the 
software (observation ID 0690744401, pn-CCD = 2, RAWX = 20, RAWY = 154).
After this careful screening we were left with 4 new SSS candidates. RGB images are presented in Fig.\,\ref{fig:SSSima} which also show an example for the appearance of a bright foreground star.
A summary of the \xmm observations 
in which they were found is given in Table~\ref{tab:observations}.

We extracted the events to produce spectra and light curves from circular regions around the source 
positions and nearby blank-sky areas. Due to the higher sensitivity of the pn compared to MOS detectors 
at low energies, we used only pn data except for the timing analysis of \sssd (see Sect.~\ref{sssd}). Single- and double-pixel events (PATTERN 0--4) were selected
excluding known bad CCD pixels and columns (FLAG 0).
For the spectra we removed times of increased flaring activity when the background was above a threshold of 8 counts ks$^{-1}$ arcmin$^{-2}$ (7.0--15.0\,keV band).
EPIC spectra were re-binned to have at least 1 count per bin and C-statistic was used. The SAS tasks {\texttt arfgen} and  {\texttt rmfgen} were used to generate 
the corresponding detector response files. The X-ray spectra were analysed with the spectral fitting package 
\xspec\,12.11.0k\footnote{Available at \url{https://heasarc.gsfc.nasa.gov/xanadu/xspec/}} \citep{1996ASPC..101...17A}.
Errors are specified for 90\% confidence, unless otherwise stated.

We modelled the EPIC-pn spectra with blackbody emission attenuated by photo-electric absorption. 
For 3 of the sources a weak, but significant high-energy tail is seen in the spectra.
Because the shape of this component is not well constrained (power law and bremsstrahlung both yield acceptable fits) 
we used a bremsstrahlung model as it is generally seen in the X-ray spectra of mCVs \citep[see e.g.][]{2017PASP..129f2001M}.
We fixed the temperature at 10\,keV and assumed that blackbody and bremsstrahlung components are attenuated by the same N$_{\rm H}$.
First, we assumed the sources are located in the LMC and used two column densities along the line of sight.
One accounts for the Galactic foreground with solar abundances according to \citet{2000ApJ...542..914W} and was fixed at the value obtained from \Hone measurements \citep{1990ARA&A..28..215D}\footnote{Extracted using NASA's HEASARC web interface \url{https://heasarc.gsfc.nasa.gov/cgi-bin/Tools/w3nh/w3nh.pl}}, 
the other (free in the fit) with metal abundances set to 0.5 reflects the absorption by the interstellar medium of the LMC \citep{2002A&A...396...53R} and local to the source. Second, if the source could be closer and located in the Milky Way, we used only one column density with solar abundance, allowing it as free parameter in the spectral fit. Luminosities were corrected for absorption and calculated assuming a distance of 50\,kpc in the case of LMC and 1--5\,kpc in the case of our Galaxy.

\subsection{\swift}
\label{sec:swift}

We produced long-term \swift XRT light curves from archival data in a soft energy band (0.3--1.0\,keV or 0.3--1.5\,keV, depending on the spectrum)  with time bins of 1\,day, using the online tool of the UK \swift Science Data Centre\footnote{\url{https://www.swift.ac.uk/user_objects/}}, which is described in \citet{2007A&A...469..379E} and \citet{2009MNRAS.397.1177E}. This time binning ensures that  observations (also with different obsids) performed shortly after each other are merged. We set the minimum fractional exposure to 0.005 to ignore exposures too short to allow the sources to be detected.
The list of available observations can also be obtained from the data centre\footnote{\url{https://www.swift.ac.uk/swift_live/}}.
Five archival observations cover the position of \sssa (\ssa for short), 3 of them have sufficient XRT exposure to detect the source (February 2012, September 2016 and June 2020). 
The XRT 0.3--1.0\,keV count rates during these observations were (1.4$^{+1.5}_{-0.9}$)\expo{-3} \ct, (2.5$^{+2.7}_{-1.6}$)\expo{-3} \ct and (0.62$^{+1.2}_{-0.59}$)\expo{-3} \ct, respectively.
Also 5 observations cover the position of \sssb (\ssb), one of them with an XRT exposure of nearly 2\,ks (March 2013). The 0.3--1.5\,keV XRT count rate during this observation was (4.4$^{+2.1}_{-1.6}$)\expo{-3} \ct. 
Numerous serendipitous observations of \sssc (\ssc) exist in the \swift archive, which allowed us to investigate the long-term variability between November 2008 and June 2016.
No \swift observations cover the position of \sssd (\ssd).

We also used the \swift/UVOT on-line analysis 
tool\footnote{\url{https://www.ssdc.asi.it/mmia/index.php?mission=swiftmastr}}, 
to search for optical/UV counterparts in the UVOT data. 
The analysis tool is based on the \swift/UVOT \citep{2005SSRv..120...95R}
software package and consists of the following steps: summing of all the exposure fractions, 
performing source detection using the  \texttt{uvotdetect} task (detection threshold of 6 
and quality filtering applied) and applying magnitude de-reddening using the E(B-V) 
value at the query position according to \citet[][]{2011ApJ...737..103S}.

\subsection{\ero}
\label{sec:eRASS}

\ssb was in the field of view of \ero, the soft X-ray instrument on the Spektrum-Roentgen-Gamma (\srg) mission \citep{2021A&A...647A...1P} during the ``CalPV phase'' \citep{2021arXiv210811517H}. 
To analyse the CalPV data we used the \ero Standard Analysis Software System \citep[\eSASS version {\tt eSASSusers\_201009,}][]{2021arXiv210614517B}. 
The data reduction and spectral analysis were performed in a similar way as described by \citet{2021arXiv210614532M} and \citet{2021arXiv210614536C}, where details can be found.
The details of the CalPV observations are given in Table~\ref{tab:observations-ero}.

\section{OGLE observations}
\label{sec:ogle}

\begin{figure}
 \resizebox{0.495\hsize}{!}{\includegraphics[clip=]{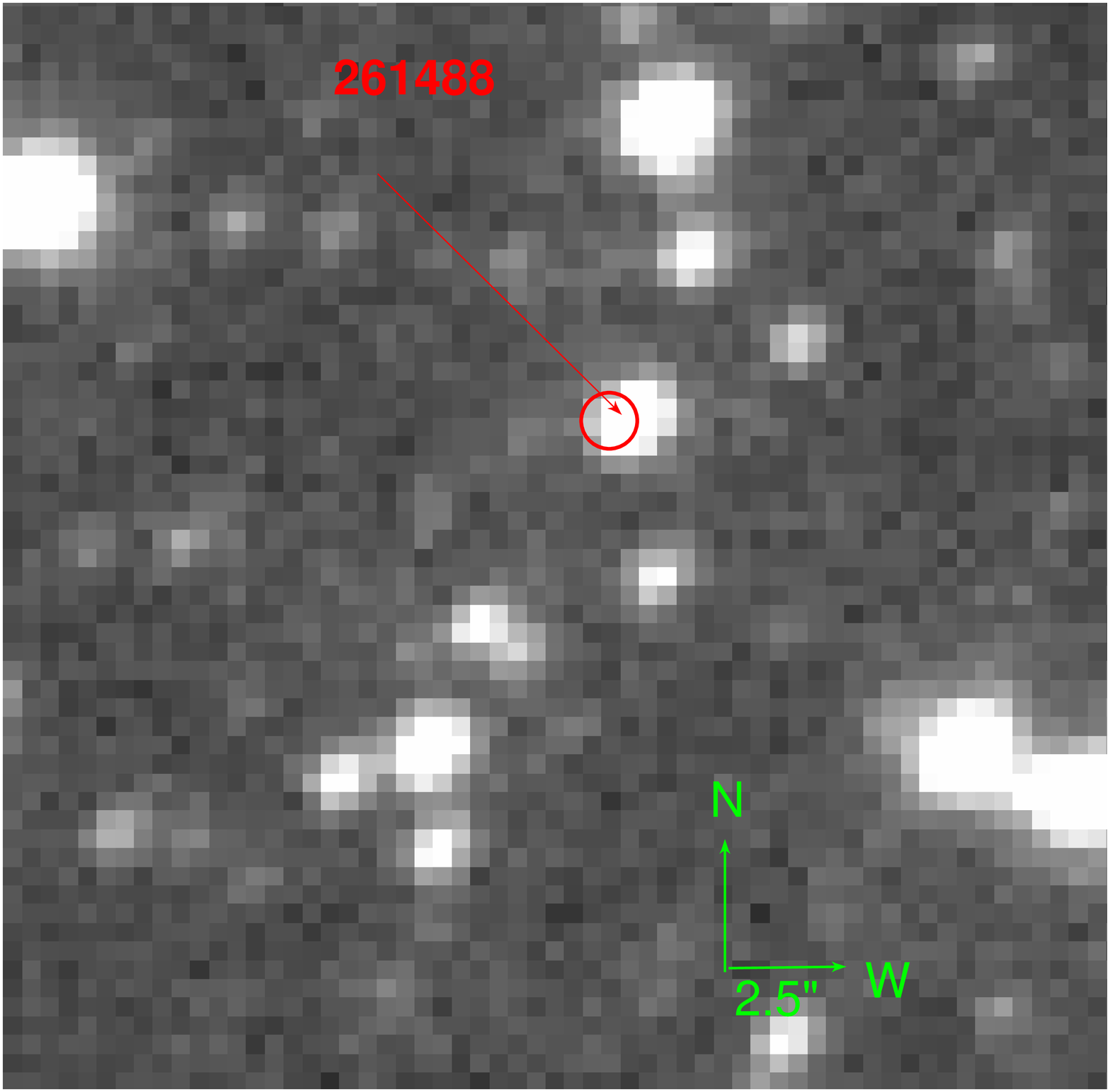}}
 \resizebox{0.495\hsize}{!}{\includegraphics[clip=]{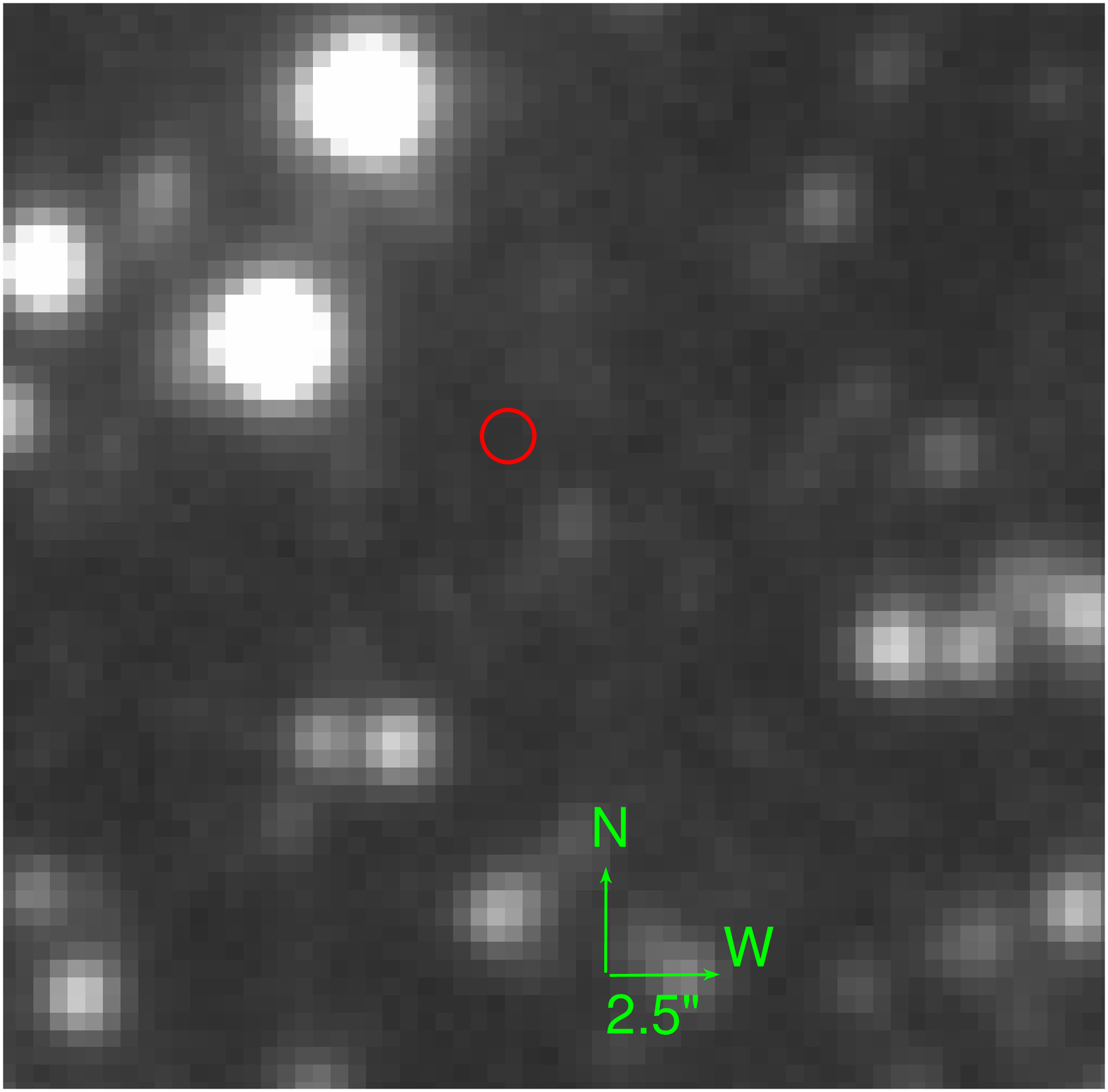}}\\
 \resizebox{0.495\hsize}{!}{\includegraphics[clip=]{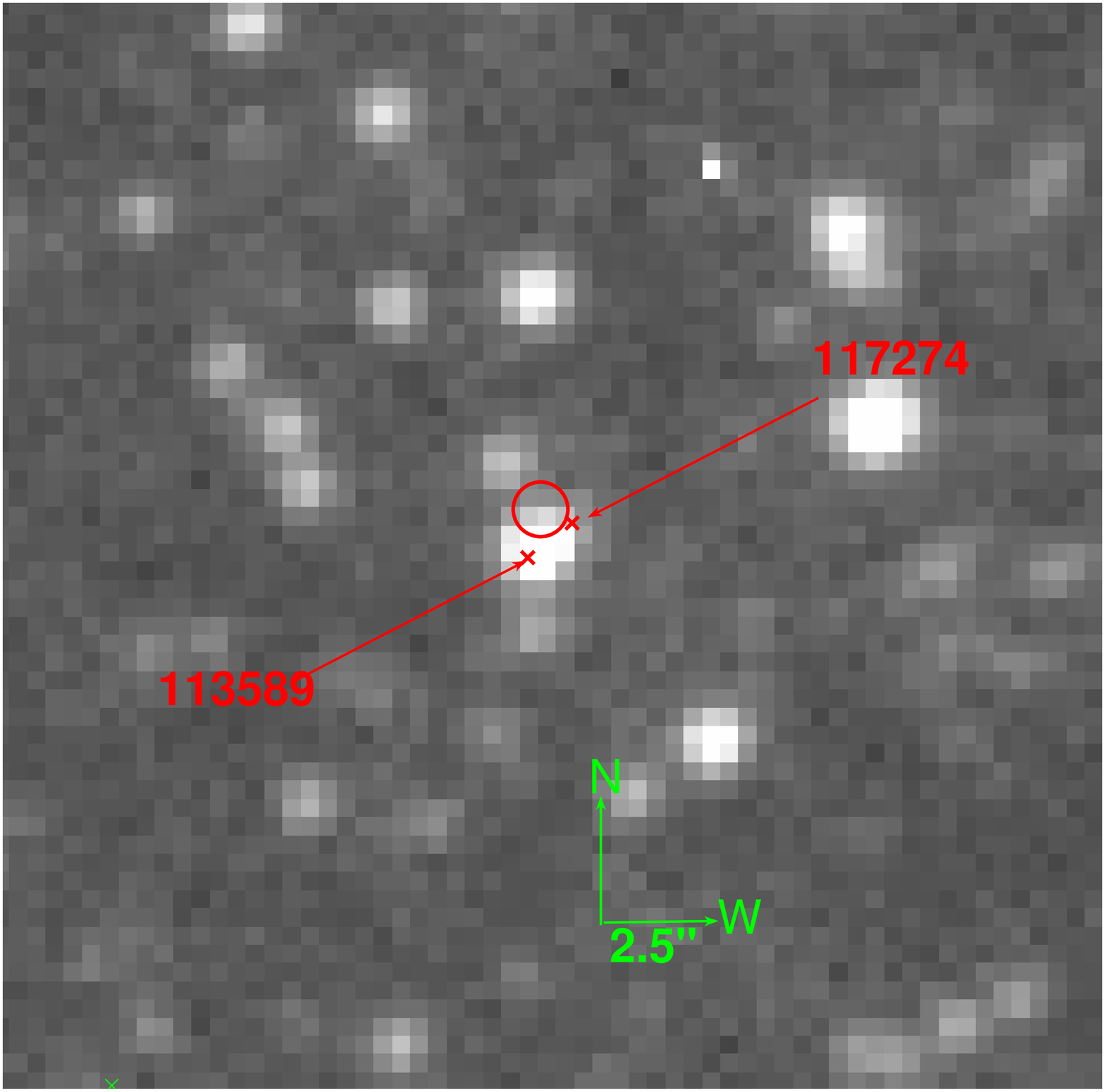}}
 \resizebox{0.495\hsize}{!}{\includegraphics[clip=]{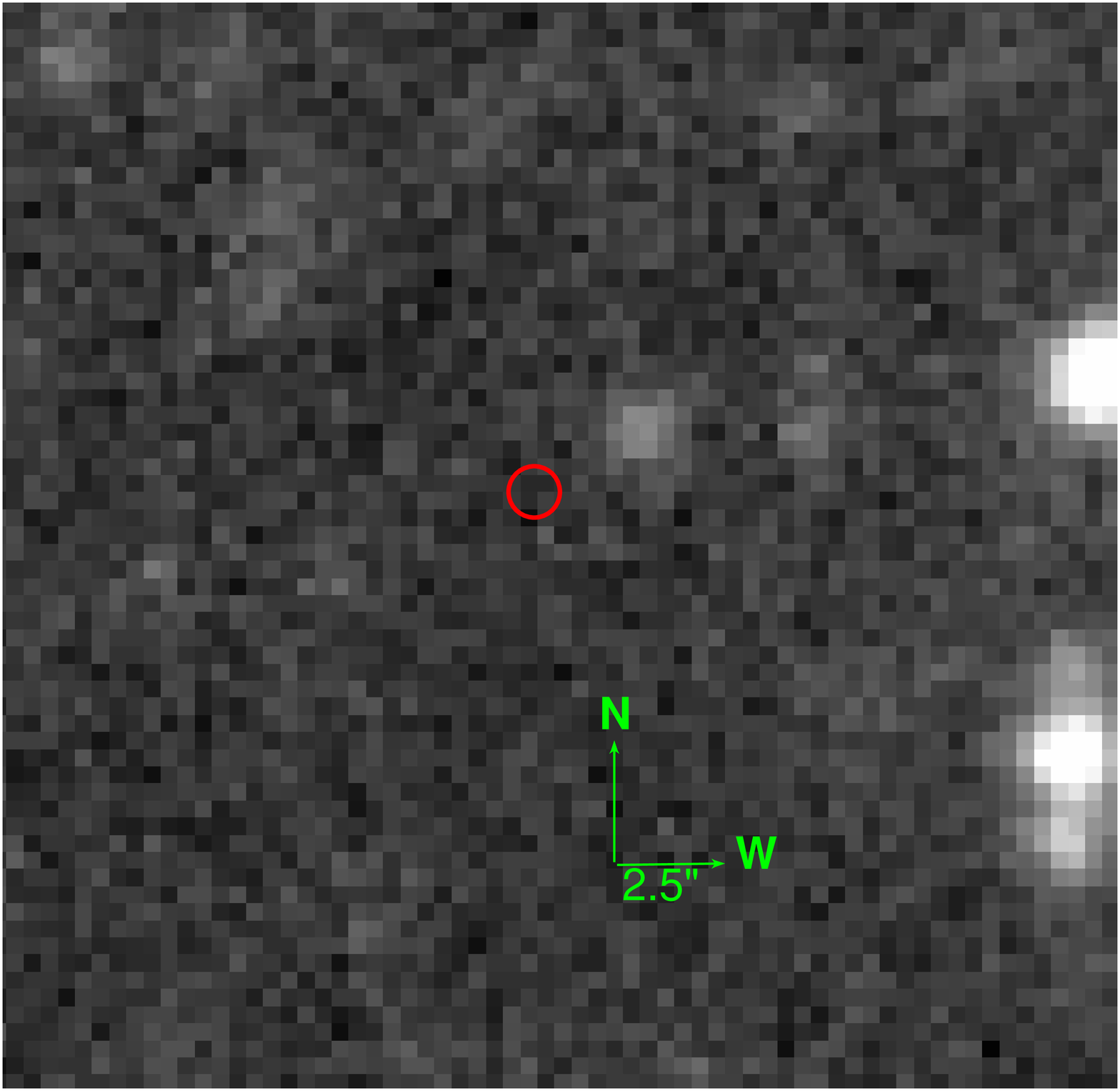}}
 \caption{Finding charts of the 4 new SSSs. On the zoomed in I-band images from OGLE-II/III, the 1$\sigma$ error circles of the \xmm positions and the OGLE source ids are marked in red. The field identifications for \ssa (top-left), \ssb (top-right)), \ssc (bottom-left), and \ssd (bottom-right) are OGLE-II LMC\_SC14, OGLE-III LMC118.6, OGLE-II LMC\_SC11 and OGLE-III LMC142.7, respectively.
 }
 \label{fig:SSSimaogle}
\end{figure}

The fields in the LMC were also observed by the Optical Gravitational Lensing Experiment \citep[OGLE;][]{1992AcA....42..253U} which provides more than twenty years of monitoring data \citep[OGLE IV;][]{2015AcA....65....1U}. We looked at images taken in the V and I filter pass-bands where the photometric magnitudes are calibrated to the standard VI system. For each source studied in this work, we searched for possible optical counterparts within the X-ray error circle given in Table~\ref{tab:observations} 
(see  Fig.~\ref{fig:SSSimaogle}).

\section{Results}
\label{sec:results}

\begin{figure*}
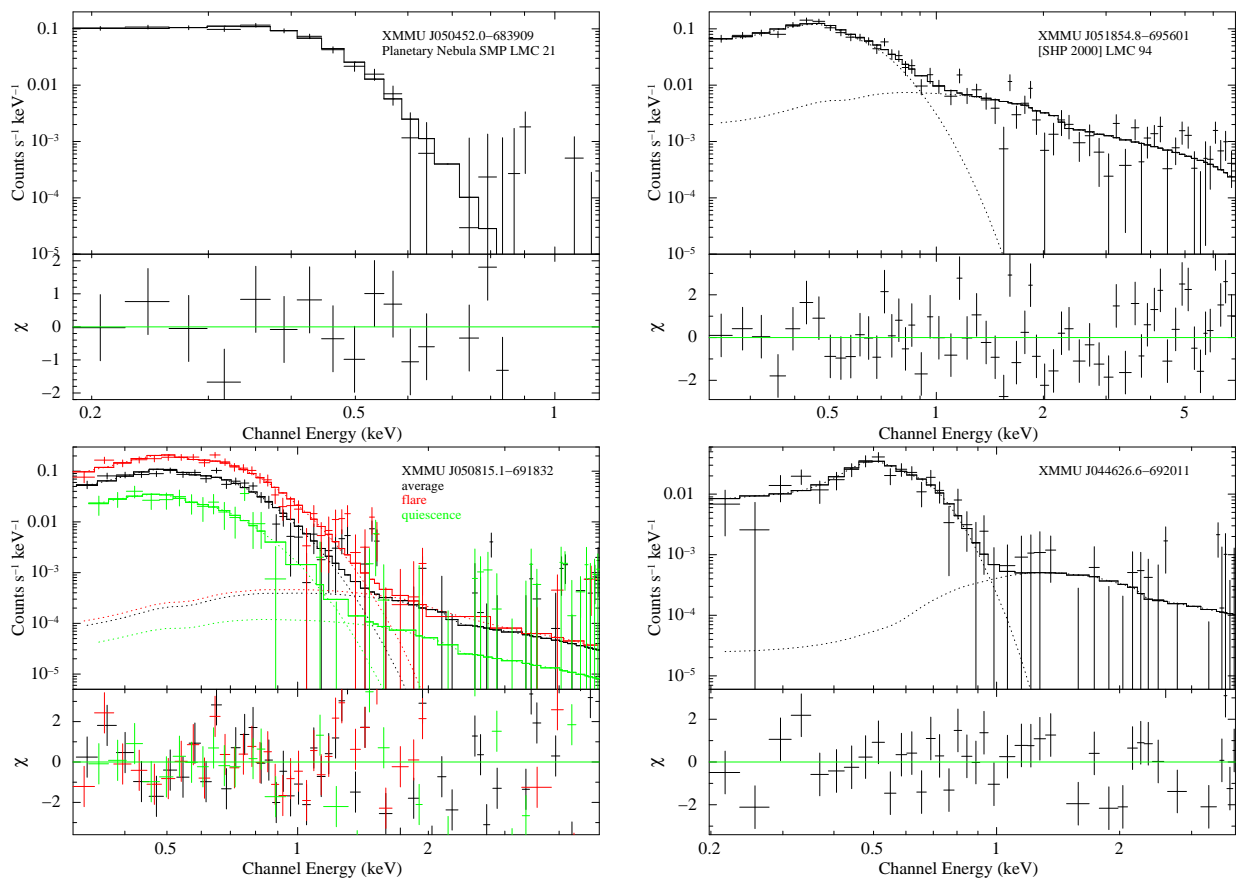

 \begin{center}
   \resizebox{0.9\hsize}{!}{\includegraphics[angle=-90,clip=]{model_bb_cstat_J0504.ps}
                           \hspace{10pt}
   \includegraphics[angle=-90,clip=]{model_bb_brems_j0518_cstat.ps}} 
   
  \resizebox{0.9\hsize}{!}{\includegraphics[angle=-90,clip=]{model_bb_brems_j0508.ps}
                           \hspace{10pt}
     \includegraphics[angle=-90,clip=]{model_bb_brems_j0446_cstat.ps}}                   
 \end{center}
  \caption{EPIC-pn X-ray spectra of new SSS observed in the \xmm survey of the LMC. 
           For J0508 the best-fit model was derived from the average spectrum, 
           while the models for quiescent and flare emission were scaled by a constant 
           factor keeping all other parameters fixed at the values of the average spectrum.}
  \label{fig:pnspectra}
\end{figure*}

\subsection{\sssa }

The SSS candidate \ssa with the softest X-ray spectrum in our sample was covered by two \xmm observations 
(Table~\ref{tab:observations}). 
It was detected in both observations in the lowest energy band (0.2--0.5\,keV) in pn and 
MOS images, clearly excluding an instrumental origin. 
The EPIC-pn count rates in the soft band (corrected for vignetting) determined by the 
source detection algorithm of (4.11$\pm$0.36)\expo{-2} \ct (February 2013) and 0.0455(14) \ct (October 2017) are consistent 
within their (1\,$\sigma$) errors. The source was observed at a very large off-axis angle in 2013. To avoid 
systematic uncertainties, we consider only the 2017 observation for the extraction of source products. 

To search for an optical counterpart, we used the X-ray coordinates derived from the observation with smaller 
off-axis angle. This position is right at the centre of the planetary nebula \smplmcn 
\citep[at R.A. = 05$^{\rm h}$04$^{\rm m}$51\fs99 and Dec. = $-$68\degr39\arcmin09\farcs7 with angular extent of 0.23\arcmin,][catalogue entry 1797]{2006MNRAS.373..521R}.
This is the second association of an SSS with a planetary nebula in the LMC after \xmmlmco = \smplmco 
\citep{2008A&A...482..237K}, while 2 are detected in the SMC \citep{2010A&A...519A..42M}. 
The OGLE counterpart of this object (LMC\_SC14 StarID: 261488) is at R.A. = 05$^{\rm h}$04$^{\rm m}$52\fs05 and Dec. = $-$68\degr39\arcmin09\farcs9 with $I=16.9$ mag (Fig.~\ref{fig:SSSimaogle}). The OGLE-II I-band light curve of the star does not exhibit significant variability. 

The EPIC-pn spectrum is well represented by an absorbed blackbody component (Fig.~\ref{fig:pnspectra} and Table~\ref{tab:spectral}).
The inferred size of the emission area is consistent with emission from the full surface of a WD (see Table \ref{tab:spectral}).
The source intrinsic luminosity of 1.7\ergs{37} is very similar to that of \smpsmc \citep{2010A&A...519A..42M} 
and consistent with being constant between the two \xmm observations, almost 4.7 years apart.
On the other hand the source was not detected in \rosat pointed observations. Using the best-fit model parameters 
derived from the EPIC-pn spectrum (obtained from the observation in 2017) and the \rosat PSPC detector response, a count rate of 3.5\expo{-3} \ct is expected.
With such a count rate the source should have been detected in two \rosat PSPC observations: 500037p (April 1992, 6826\,s exposure)
and 500258p (October -- December 1993, 12715\,s) with $\sim$20\,cts (taking into account vignetting), which is a factor of about 3 higher 
than the detection limit. I.e. \ssa was at least a factor of 3 fainter during the \rosat observations.
During \swift observations (2012, 2016 and 2020) \ssa was detected with 0.3--1.0\,keV count rates between 0.6\expo{-3} \ct and 2.5\expo{-3} \ct. 
Assuming the EPIC-pn best-fit spectral parameters, an XRT count rate of $\sim$2.1\expo{-3} \ct is expected, consistent with the observations.

\begin{table*}
\centering
\caption[]{Spectral fit results.}
\begin{tabular}{llcccccccc}
\hline\hline\noalign{\smallskip}
\multicolumn{1}{c}{Source} &
\multicolumn{1}{c}{Observation} &
\multicolumn{1}{c}{kT$_{\rm bb}$} &
\multicolumn{1}{c}{N$_{\rm H}^{\rm Gal}$} &
\multicolumn{1}{c}{N$_{\rm H}^{\rm LMC}$} &
\multicolumn{1}{c}{C-stat} &
\multicolumn{1}{c}{dof} &
\multicolumn{1}{c}{F$_{\rm observed}$\tablefootmark{c}} &
\multicolumn{1}{c}{L\tablefootmark{d}} &
\multicolumn{1}{c}{R$_{\rm BB}$} \\
\multicolumn{1}{c}{name\tablefootmark{a}} &
\multicolumn{1}{c}{ID} &
\multicolumn{1}{c}{(eV)} &
\multicolumn{1}{c}{(\ohcm{21})} &
\multicolumn{1}{c}{(\ohcm{21})} &
\multicolumn{1}{c}{/dof} &
\multicolumn{1}{c}{} &
\multicolumn{1}{c}{(erg cm$^{-2}$ s$^{-1}$)}&
\multicolumn{1}{c}{(erg s$^{-1}$)} &
\multicolumn{1}{c}{(km)} \\
\noalign{\smallskip}\hline\noalign{\smallskip}
  \ssa & 0803460101 & 31 $\pm$  4 & 0.64  & 1.2$^{+0.5}_{-0.4}$   & 0.66 & 19 & $7.6\pm0.4 \times 10^{-14}$ &  $1.7 \times 10^{37}$ & 16000$^{+40600}_{-9600}$ \\
  \noalign{\smallskip}
  \ssb & 0690751801 & 78 $\pm$  8 & 0.67  & 0.9$^{+0.6}_{-0.5}$   & 1.2 & 94 & $1.0\pm0.1 \times 10^{-13}$ &  $2.1 \times 10^{35}$ & 220$^{+140}_{-80}$ \\
  \noalign{\smallskip}
       & \ero       & 88 $\pm$  4 & 0.67  & 0.09$^{+0.16}_{-0.09}$ & 1.1  & 106 & $1.7\pm0.2 \times 10^{-13}$ &  $2.0 \times 10^{35}$ & 135$^{+28}_{-20}$ \\
  \noalign{\smallskip}
  \ssc\tablefootmark{b} & 0690752001 & 86 $\pm$  9 & 0.74  & 1.9 $^{+0.9}_{-0.8}$ & 1.7 & 54 & $1.7\pm0.1 \times 10^{-13}$ &  $5.7 \times 10^{35}$ & 290 $^{+280}_{-120}$ \\
  \noalign{\smallskip}
  \ssd & 0801990301 & 49 $\pm$ 14 & 0.72  & 6.4$^{+5.6}_{-3.2}$   & 0.94 & 40 & $2.0\pm0.3 \times 10^{-14}$ &  $7 \times 10^{36}$ & 3400$^{+9600}_{-3000}$ \\
\noalign{\smallskip}\hline\noalign{\smallskip}
  \ssb  & 0690751801 & 78 $\pm$  7 & 1.6$^{+0.5}_{-0.4}$  & --    & 1.2 & 94 & $1.0 \times 10^{-13}$ &  $9.0 \times 10^{31}$ &  4.5 $^{+2.9}_{-1.6}$ \\
  \noalign{\smallskip}
        & \ero       & 88 $\pm$  4 & 0.75$^{+0.15}_{-0.13}$  & -- & 1.1 & 106 & $1.7 \times 10^{-13}$ &  $6.0 \times 10^{31}$ &  2.7 $^{+0.6}_{-0.5}$ \\
  \noalign{\smallskip}
  \ssc\tablefootmark{b} & 0690752001 & 87 $\pm$  9 & 2.5$^{+1.0}_{-0.8}$  & --    & 1.7 & 54 & $1.7 \times 10^{-13}$ &  $2.4 \times 10^{32}$ & 6.1$^{+6.7}_{-2.8}$ \\
  \noalign{\smallskip}
  \ssd & 0801990301 & 50 $\pm$ 14 & 7.1$^{+6.5}_{-3.3}$ & --    & 0.96 & 40 & $2.0 \times 10^{-14}$ & $5.0 \times 10^{33}$ & 87$^{+265}_{-78}$ \\
\noalign{\smallskip}\hline

\end{tabular}
\tablefoot{
Best-fit parameters using a model with absorbed blackbody and bremsstrahlung emission. The bremsstrahlung component is not well constrained and the temperature was fixed at 10\,keV (see Sect.\,\ref{sec:XMM}). The C-stat values include the full model.
Errors indicate 90\% confidence ranges.
\tablefoottext{a}{For full source names see Table\,\ref{tab:observations}.}
\tablefoottext{b}{Spectral parameters derived from average spectrum.}
\tablefoottext{c}{Fluxes are provided for the 0.1--2.4\,keV band to allow comparison with values published for SSSs based on \rosat observations.}
\tablefoottext{d}{Source luminosities (0.1--2.4\,keV) corrected for absorption, assuming a distance of 50\,kpc \citep{2013Natur.495...76P} in the upper part of the table. The Galactic foreground column density was taken from \citet{1990ARA&A..28..215D}.
For the model in the lower part of the table only a Galactic absorption component (with free column density in the fit) and a distance of 1\,kpc were assumed.
A note of caution to absorption corrected X-ray luminosities in soft energy bands: Column densities with large error lead to a large uncertainty on the luminosity, which is in particular the case for \ssd.}
}
\label{tab:spectral}
\end{table*}

\subsection{\sssb}

\ssb was detected in the EPIC images of observation 0690751801 (Table~\ref{tab:observations}) with an EPIC-pn count rate (0.2--0.5\,keV, corrected for vignetting) of (4.29$\pm$0.20)\expo{-2} \ct. No optical counterpart was found for the object in the OGLE database (Fig.~\ref{fig:SSSimaogle}) nor in the \gaia catalogues within the X-ray error circle. 
A star with $V\sim21$ mag was however found within the error circle from the Magellanic Clouds Photometric Survey (MCPS) catalogue
of \cite{2004AJ....128.1606Z}. No source was detected at the position of \ssb from the \swift/UVOT data. Using the longest UVOT exposure covering \ssb (obsid 00045502001, exposure 1014\,s, filter uvm2) and the sources detected in the field, a lower limit of 18.5 mag was obtained.

Due to the relatively small off-axis angle of the source and the high detection significance the data were suitable for a detailed analysis. 
The EPIC-pn spectrum can be fit by a blackbody and an additional bremsstrahlung component to account for a significant hard tail in the spectrum. 
Figure\,\ref{fig:pnspectra} shows the best-fit EPIC-pn spectrum, and the best-fit spectral parameters are tabulated in Table~\ref{tab:spectral}. 
The total 0.1--8.0\,keV absorption-corrected luminosity at the distance of the LMC is 2.13\ergs{35} with the bremsstrahlung component accounting for 11\% demonstrating the dominance of the soft component in the spectrum. 
Although the blackbody luminosity depends somewhat on the temperature and absorption derived from the spectral fit, the luminosity is well below the range in which H-burning is expected to be stable \citep{2013ApJ...777..136W}.
In addition, the radius of the emitting area of 220 km would be much smaller than the size of a WD.
On the other hand, the X-ray luminosity would be too high to be explained by an mCV in the LMC. If we assume the source to be located in the Milky Way, luminosity and radius of the emitting area are correspondingly smaller (e.g. for distances of 1 to 5 kpc: \lbol\ = 9\ergs{31} to 2\ergs{33} and R$_{\rm BB}$ = 4 to 20 km, see also Table\,\ref{tab:spectral}).

\ssb was observed during the CalPV phase of \ero. Data from the different telescope modules TM1, 2, 3, 4, and 6 (which are equipped with cameras with on-chip optical blocking filter) and from the 3 observations were added. The combined spectrum is well represented by the same models as in the case of \xmm EPIC-pn, the best-fit parameters are listed in Table\,\ref{tab:spectral}. 
 
\begin{figure}
 \begin{center}
   \resizebox{0.95\hsize}{!}{\includegraphics[angle=-90,clip=]{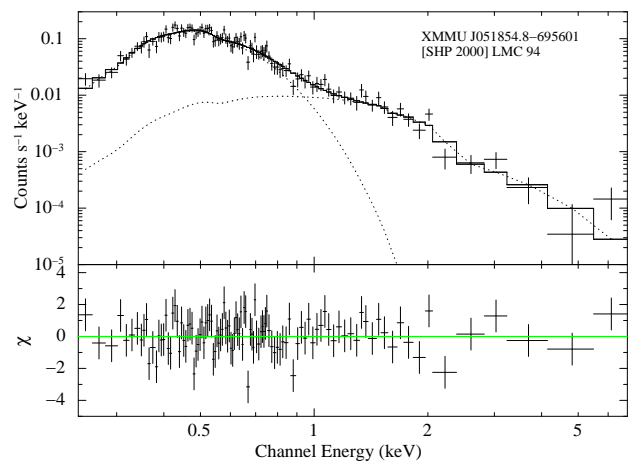}}
 \end{center}
 \caption{Combined \ero X-ray spectrum of \sssb observed in the CalPV phase together with the 
          best-fit model consisting of soft blackbody and hard bremsstrahlung components, 
          which are indicated by dotted lines.}
 \label{fig:erospectra}
\end{figure}

The \ero spectrum with the best-fit two-absorption component model is presented in Fig.\,\ref{fig:erospectra}. 
While temperature and emission radius derived from the \ero spectrum are consistent with those measured from EPIC-pn, the column densities are significantly lower during the \ero CalPV phase in November 2019, almost seven years after the \xmm observation.
The higher absorption in 2012 results in a lower observed flux, but the absorption-corrected intrinsic luminosities during the two observations are similar, with a slightly lower value during the 2019 observations.

The source was detected previously in a \rosat HRI observation (RH601040 in December 1997/January 1998) and is listed in the catalogue of \citet{2000A&AS..143..391S} as source 94 ([SHP2000]\,LMC\,94) with  a count rate of (2.6$\pm$0.4)\expo{-3} \ct. Assuming the current spectral model derived from EPIC-pn, the expected HRI count rate is $\sim$2.2\expo{-3} \ct, consistent with the \rosat measurement. Similarly, the flux of 1.0\ergcm{-13} observed with EPIC-pn (in September 2012) corresponds to a \swift XRT 0.3--1.5\,keV count rate of $\sim$4.1\expo{-3} \ct, consistent with the detection in March 2013.

\subsection{\sssc}

\ssc was detected in the EPIC images of observation 0690752001 (Table~\ref{tab:observations}) at an EPIC-pn count rate (0.2--0.5\,keV, corrected for vignetting) of (6.25$\pm$0.30)\expo{-2} \ct. 
Although there was no OGLE counterpart within the X-ray error circle, the nearest counterpart LMC\_SC11 StarID: 117274 lies at 0.75\arcsec\ (just outside the error circle) with I = 20.2\,mag (Fig.~\ref{fig:SSSimaogle}). 
Due to the detection of the source near the edge of the field of view, the position may be uncertain and we ascertain this as the plausible optical counterpart.
The OGLE-II I-band light curve of the star does not exhibit variability although this cannot be ruled out due to large uncertainties. No source was detected at the position of \ssc from the \swift/UVOT data. Using the longest UVOT exposure covering \ssb (Obsid 00030348002, exposure 2049\,s, filter w1) and the sources detected in the field, a lower limit of 19.1 mag was obtained.

\begin{figure}
  
  \resizebox{\hsize}{!}{\includegraphics[angle=-90]{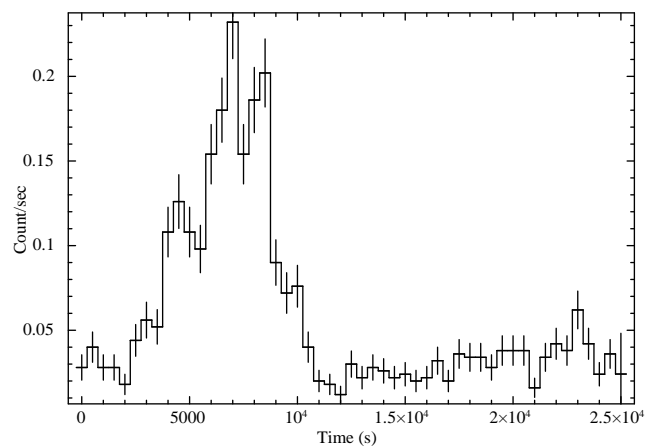}}
  \caption{
    EPIC-pn light curve of \sssc (observation 0690752001) binned at 500\,s in the 0.2--2\,keV band.
    Time zero corresponds to 2012-09-22 at 02:29:24 UTC.
  }
  \label{fig:sssc_lc}
\end{figure}

Although the source was detected at a large off-axis angle, its relatively bright nature allowed us to perform timing and spectral analysis of the EPIC-pn data. 
The EPIC-pn light curve of \ssc in the 0.2--2\,keV band shows flaring activity shortly after the beginning of the observation lasting for $\sim$8.2\,ks (Fig.~\ref{fig:sssc_lc}). 

The EPIC-pn spectrum can be described by blackbody emission with $kT=86$\,eV and a weak bremsstrahlung component. 
The best-fit spectral parameters are summarised in Table~\ref{tab:spectral}. 
The blackbody temperature of \ssc is similar to that of \ssb, but significantly higher than that of \ssa and probably also \ssd.
The total unabsorbed luminosity at the distance of LMC is 6\ergs{35} with the bremsstrahlung component accounting for  only 2\% of the total 0.1--8\,keV emission demonstrating again the dominance of the soft component. As in \ssb\, the luminosity is well below the range for stable H-burning and the radius of the emitting area of $\sim$290\,km would be much smaller than the size of a WD.
The inferred size of the emission region is indicative for hot-spots on the WD surface and suggests an IP nature of the source. 
However, the X-ray luminosity would again be too high to be explained by an LMC membership of the source. If we assume a source location in the Milky Way, luminosity and radius of the emitting area are correspondingly smaller (e.g. for distances of 1 to 5 kpc: \lbol\ = 2\ergs{32} to 5\ergs{33} and R$_{\rm BB}$ = 6 to 30 km, Table\,\ref{tab:spectral}).

We also extracted spectra as shown in Fig.~\ref{fig:pnspectra}, from the `flaring` and the `quiescent` intervals guided by the light curve in Fig.~\ref{fig:sssc_lc}. The spectrum of the entire observation was fit simultaneously together with the spectra from the flaring and the quiescent intervals with the blackbody temperature, the N$_{\rm H}^{\rm LMC}$ and a  normalisation constant (to account for an overall change in flux) left free. 
Due to the insufficient statistical quality of the spectra no change in the spectral parameters except the normalisation could be detected within uncertainties. The peak luminosity during the flaring interval was a factor of $\sim$10 higher than during quiescence. 
Figure\,\ref{fig:pnspectra} shows the best-fit EPIC-pn spectra during the entire observation, the flaring and quiescence intervals.

\ssc is consistent in position with 2RXS\,J050813.5$-$691831, which was detected in the \rosat all-sky survey \citep{2016A&A...588A.103B} with a PSPC count rate of 1.2\expo{-2} \ct and the source is also listed in the \rosat HRI catalogue of \citet{2000A&AS..143..391S} with a count rate of (4.9$\pm$0.7)\expo{-3} \ct ([SHP2000]\,LMC\,38). 
Using the best-fit model parameters derived from the EPIC-pn spectrum and the \rosat PSPC and HRI detector responses, 
the predicted count rates are $\sim$1.2\expo{-2} \ct and $\sim$4.5\expo{-3} \ct, respectively. This indicates that the source was at a consistent brightness level during the \rosat all-sky survey (1990), the pointed \rosat HRI (RH601036 1997/1998) and the \xmm (2012) observation. 

\begin{figure}
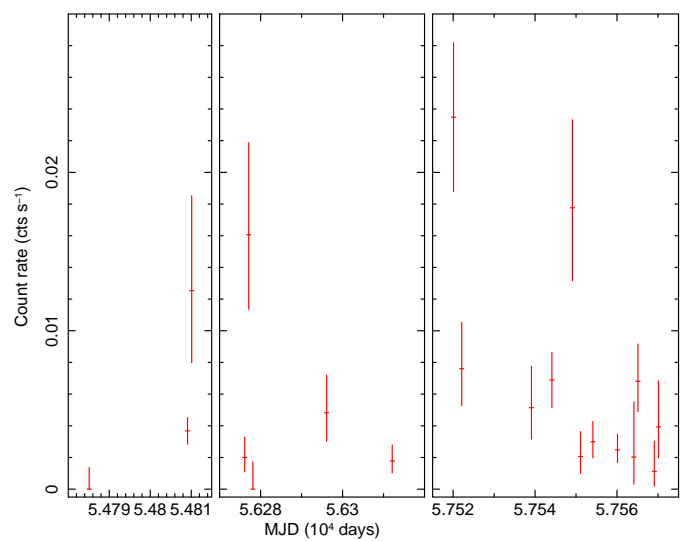

  \resizebox{0.98\hsize}{!}{
  \includegraphics[clip=,angle=-90]{Swift_J0508_LCepoch1.ps}
  \includegraphics[clip=,angle=-90]{Swift_J0508_LCepoch2.ps}
  \includegraphics[clip=,angle=-90]{Swift_J0508_LCepoch3.ps}}
  \caption{
    Long-term X-ray light curve of \sssc in the 0.3--1.5\,keV band with one-day binning obtained from archival \swift observations.
    The source was mainly observed during three epochs: November/December 2008, December 2012/January 2013 and May/June 2016.
  }
  \label{fig:sssc_swift}
\end{figure}

The soft X-ray light curve of \ssc extracted from archival \swift data (Fig.\,\ref{fig:sssc_swift}) suggests some brightening during May/June 2016.
Based on the EPIC-pn best-fit spectral model parameters, the observed flux of 1.7\ergcm{-13} corresponds to a \swift XRT count rate of $\sim$7.4\expo{-3} \ct (0.3--1.5\,keV). 
During June 2016 the source flux reached a maximum about a factor of 3 higher while the average count rate from the full \swift light curve of (3.1$\pm$0.4)\expo{-3} \ct is at about half the flux level during the \rosat and \xmm observations.

\subsection{\sssd}
\label{sssd}

\ssd was detected in the EPIC images observation 0801990301 (Table~\ref{tab:observations}) at an EPIC-pn count rate (0.2--0.5\,keV, corrected for vignetting) of (8.9$\pm$0.8)\expo{-3} \ct.  No optical counterpart was found for the object in the OGLE database (Fig.~\ref{fig:SSSimaogle}) nor in the \gaia catalogues within the X-ray error circle. 

\begin{figure}
 \resizebox{\hsize}{!}{\includegraphics[clip=]{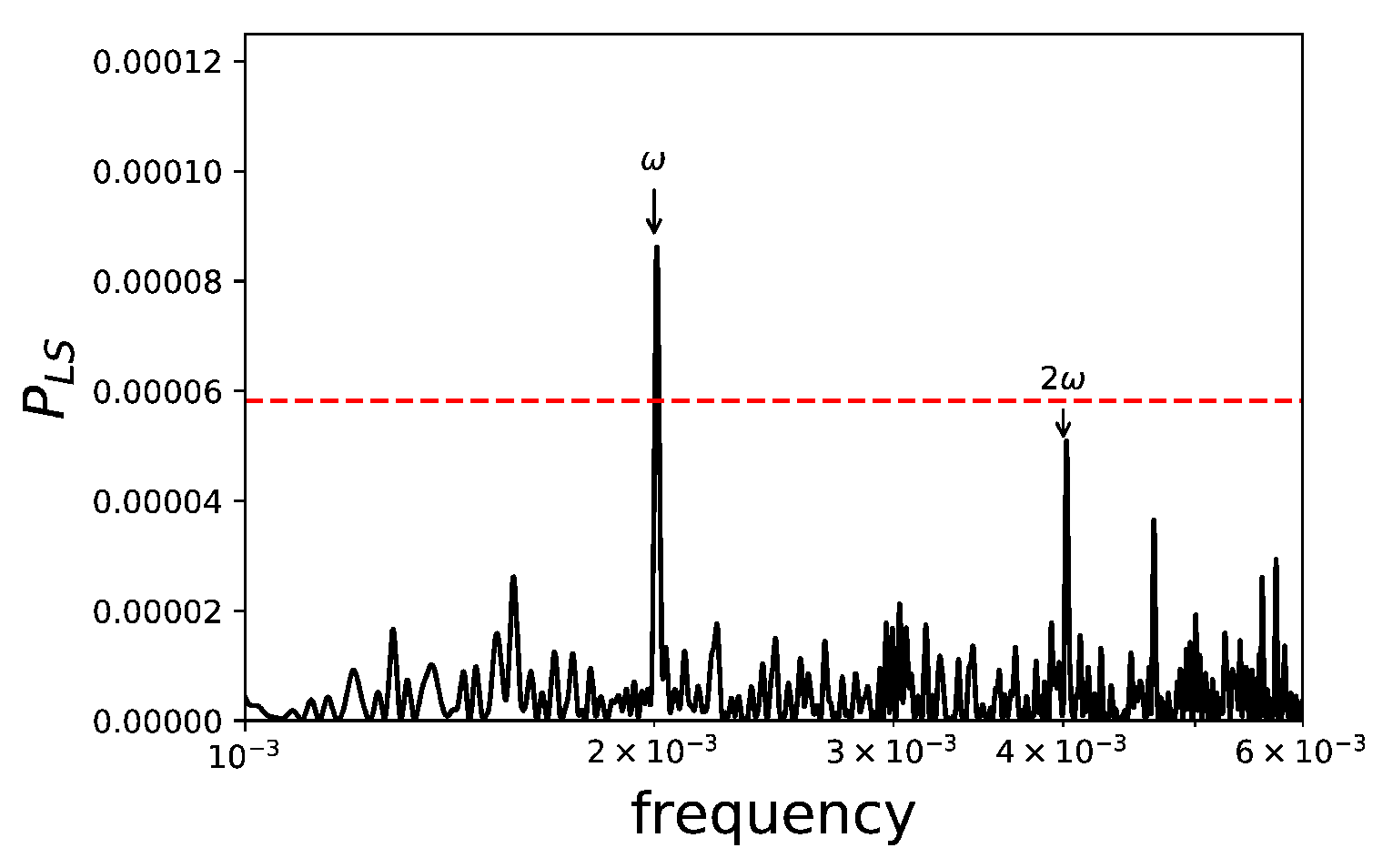}}
 \caption{
   Lomb-Scargle periodogram of the combined EPIC light curve of \sssd (0.2--2\,keV). The dashed red line marks the 3$\sigma$ confidence level.
 }
 \label{fig:LS}
\end{figure}

\begin{figure}
  \resizebox{0.9\hsize}{!}{\includegraphics[angle=-90,clip=]{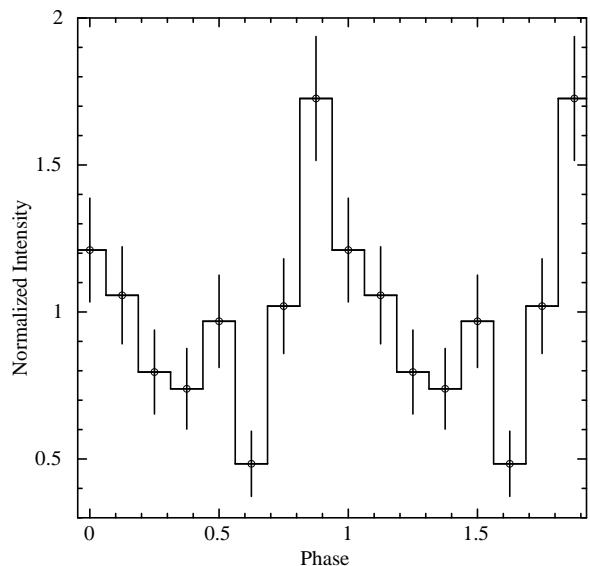}}
  \caption{
    EPIC-pn light curve of \sssd in the 0.2--2\,keV band folded at the best-fit spin period.
  }
  \label{fig:s_foldlc}
\end{figure}

The source was detected at a relatively small off-axis angle, therefore allowing detailed timing and spectral analysis of the EPIC-pn data.
We first analysed the EPIC-pn light curve in the energy range between 0.2--2\,keV using a Lomb-Scargle periodogram analysis \citep{1976Ap&SS..39..447L,1982ApJ...263..835S}. A period was found at $\sim$496\,s along with the first harmonic. The signal was somewhat enhanced when using the combined EPIC (pn+M1+M2) light curve and the  periodogram is shown in Fig.~\ref{fig:LS}.
The pulse period and its associated 1$\sigma$ error are determined to  497.4$\pm$1.6\,s.
The \xmm EPIC-pn light curve in the range of 0.2--2\,keV, folded with the best-obtained period is shown in Fig.~\ref{fig:s_foldlc}.
The period is in the typical range observed from intermediate polars and likely indicates the spin period of the WD.

The EPIC-pn spectrum is well represented by an absorbed blackbody component with an indication for a weak hard tail (Fig.~\ref{fig:pnspectra}). The the best-fit spectral parameters are listed in Table\,\ref{tab:spectral}.
Because of very large uncertainties in temperature and absorption, the luminosity of the blackbody component can not be reliably determined.
Assuming the formally best fit parameters, the contribution of the bremsstrahlung emission to the total luminosity is completely negligible and could also be caused by a background-subtraction problem.
Assuming \ssd is located at LMC distance, the inferred size of the emission area is consistent with emission from the full surface of a WD. The measured LMC \nh=6.4$\times$\ohcm{21} is highest among the sources studied in this work and leads to an unabsorbed luminosity of 7\ergs{36}. We caution that this value is strongly dependent on the actual absorption along the line of sight to this source. If we assume a source location in the Milky Way at a distance of 1\,kpc, luminosity and radius of the emitting area are \lbol\ = 5\ergs{33} and R$_{\rm BB}$ = 87 km (Table\,\ref{tab:spectral}).

\ssd is located at the western rim of the LMC and was only covered by a short (9\,ks) \rosat HRI observation with the source at a large off-axis angle of $\sim$17\arcmin, which does not yield a stringent upper limit for the flux. No pointed \rosat observations with the more sensitive PSPC detector, nor \swift observations were performed at that location. 

\section{Discussion}
\label{sec:discussion}

From a systematic search for new SSSs in the \xmm data of the Magellanic Clouds we report the discovery of 4 sources in the direction of the LMC. We also used OGLE data to search for their possible optical counterparts and used archival \rosat and \swift observations to investigate their long-term behaviour. One of the sources, \sssb was in the field of view of \ero during the CalPV phase and was studied in this work.

\subsection{\sssa = \smplmcn}
\ssa, is identified as central star of the planetary nebula (PN) \smplmcn in the LMC. \smplmcn
is a high excitation PN \citep{1984MNRAS.208..633M} exhibiting a quadrupolar morphology of 1.15\arcsec\ extent \citep{1999ApJ...510..687S}. The chemical composition from optical observations suggest a Type I PN hosting a relatively massive AGB progenitor \citep[][and references therein]{2007ApJ...671.1669S}.
Using archival observations from 2009,  \cite{2010A&A...519A..42M} derived an upper limit of the unabsorbed X-ray luminosity to be 3\ergs{35} (see Table\,5 in their paper). In this work we detect the source with high confidence, and our results indicate a much higher X-ray luminosity.
The soft spectrum and derived high luminosity ($\sim$2\ergs{37}) implies that its X-ray emission originates from the central
star rather than from the surrounding hot shocked gas. 
After \smplmco \citep{2008A&A...482..237K} this is the second known planetary nebula in the LMC to emit soft X-rays. Two such systems are also known in the SMC \citep[\smpsmc and \hbox{SMP\,SMC\,25};][]{2010A&A...519A..42M}, and from a simple estimate based on the mass ratio of the two galaxies, one would expect many more to be detected in the LMC. This deficit could be caused partly by the lower coverage of the LMC with X-ray observations, but also by the stronger diffuse soft X-ray emission in the LMC, which reduces the sensitivity for the detection of faint point sources with soft X-ray spectrum.  

\subsection{\sssb and \sssc : Galactic soft intermediate polars}

\ssb and \ssc show very similar spectral properties. Their soft blackbody-like emission can be characterised by a temperature between 70--100\,eV, both show a hard tail dominating the spectrum above 2\,keV, an absorption-corrected X-ray luminosity of a few \oergs{35} (assuming LMC distance) and a corresponding size of the emission area of $\sim$100--400\,km in radius. 
All of the above is reminiscent to the properties of soft IPs, except that the luminosities and emission areas are larger than those measured from soft IPs in the Galaxy (for a comparison see Table\,\ref{tab:softIP}).
Alternatively, the 2 sources could be located at distances around 1\,kpc, which would make their luminosities and emission radii consistent with the typical values measured from Galactic soft IPs. For this purpose, we simplified the spectral model by fitting only one column density for gas with solar elemental abundances. The inferred best-fit parameters are listed in the lower part of Table\,\ref{tab:spectral} and are consistent with those measured for Galactic soft IPs (Table\,\ref{tab:softIP}).

\begin{table*}
\centering
\caption[]{Cataclysmic variables with blackbody-like soft X-ray emission and fast spinning white dwarfs.}
\begin{tabular}{llcccccc}
\hline\hline\noalign{\smallskip}
\multicolumn{1}{l}{Source} &
\multicolumn{1}{l}{Other name} &
\multicolumn{1}{c}{Periods\tablefootmark{a}} &
\multicolumn{1}{c}{Distance\tablefootmark{b}} &
\multicolumn{1}{c}{kT\tablefootmark{c}} &
\multicolumn{1}{c}{L$_{\rm BB}$} &
\multicolumn{1}{c}{R$_{\rm BB}$} &
\multicolumn{1}{c}{Fluxes\tablefootmark{d}} \\
\multicolumn{1}{l}{} &
\multicolumn{1}{l}{} &
\multicolumn{1}{c}{} &
\multicolumn{1}{c}{(pc)} &
\multicolumn{1}{c}{(eV)} &
\multicolumn{1}{c}{(erg s$^{-1}$)} &
\multicolumn{1}{c}{(km)} &
\multicolumn{1}{c}{(erg cm$^{-2}$ s$^{-1}$)} \\
\noalign{\smallskip}\hline\noalign{\smallskip}
 RX\,J0512.2-3241       & UU Col     & 863.5 s  / 3.45 h       & 2429  & 73        & 2.1\expo{32}  &  7.6 & 5.1\expo{-13} -- 3.5\expo{-12} \\ 
 RX\,J0558.0+5353       & V 405 Aur  & 545.5 s  / 4.15 h       &  662  & 64.8      & 2.3\expo{33}  & 31.8 & 1.0\expo{-11} -- 2.3\expo{-11} \\ 
 1E\,0830.9-2238        & WX Pyx     & 1557.3 s / $\sim$5.54 h & 1591  & 82        & 1.8\expo{32}  &  5.6 & 2.3\expo{-13} -- 4.9\expo{-13} \\ 
 RE\,0751+14            & PQ Gem     & 833.7 s  / 5.19 h       &  750  & 47.6      & 9.0\expo{32}  & 36.9 & 4.7\expo{-12} -- 6.2\expo{-11} \\ 
 1RXS\,J154814.5-452845 & NY Lup     & 693.0 s  / 9.87         & 1228  & 104       & 7.8\expo{32}  &  7.2 & 4.7\expo{-12} -- 1.8\expo{-11} \\ 

\noalign{\smallskip}\hline
\end{tabular}
\tablefoot{
\tablefoottext{a}{Spin period of the WD and binary period.}
\tablefoottext{b}{Distances from \gaia DR2 \citep{2018AJ....156...58B}.}
\tablefoottext{c}{Blackbody temperatures and bolometric fluxes to compute the bolometric luminosity L$_{\rm BB}$ and the emission radius R$_{\rm BB}$ of the blackbody component 
are taken from \citet{2007ApJ...663.1277E}.}
\tablefoottext{d}{Observed flux range in the 0.2--2\,keV band, for details see Sect.\,\ref{sec:discussion}.}
}
\label{tab:softIP}
\end{table*}

\ssb is likely a persistent SSS at least since the  1990's as the detections in archival \rosat and \swift data and the recent \ero CalPV observations suggest. A variation in the amount of absorption column density is however observed between the \xmm and \ero observations in 2012 and 2019 respectively leading to significantly different observed fluxes.
 \ssc exhibits variability on timescales of hours as seen in form of flaring behaviour in the \xmm observation. IPs are known to exhibit flaring behaviour on short timescales, which may arise in the turbulent inner region of the accretion disk, or from bright spots on the surface of the white dwarf \citep{Barbera2017AccretionDC,2001cvs..book.....H}.
The largest flux variations on long-term timescales by about a factor of 6 were also seen from \ssc (Fig.\,\ref{fig:sssc_swift}), 
that could be caused by flaring activity similar as seen during the \xmm observation (Fig.~\ref{fig:sssc_lc}).
For comparison, we used the HIgh-energy LIght-curve GeneraTor\footnote{\url{http://xmmuls.esac.esa.int/upperlimitserver/}} to produce 0.2--2\,keV light curves from archival \rosat, \xmm and \swift observations of the Galactic soft IPs that are listed in Table\,\ref{tab:softIP}. 
From the available spectral models we selected the blackbody with kT = 60\,eV and N$_{\rm H}$ = 1\hcm{21} and 
the inferred minimum and maximum fluxes are summarised in Table\,\ref{tab:softIP}.
The somewhat higher variability of the Galactic soft IPs  as seen in Table.~\ref{tab:softIP} is likely due to the larger number of available observations compared to the newly discovered systems.

\subsection{\sssd}

The X-ray spectrum and the pulsations of $\sim$497\,s suggest an mCV nature also for \ssd. The coherent pulsations denote the spin period of the white dwarf. Several other SSSs with measured spin period were discovered in \xmm observations of M\,31; 
XMMU\,J004319.4+411758 a transient source with period of $\sim$865\,s \citep{2001A&A...378..800O} and 
XMMU\,J004252.5+411540 with a period of $\sim$218\,s \citep{2008ApJ...676.1218T}.
In one case an SSS with 1110\,s pulsations could be identified with nova M31N 2007-12b \citep{2011A&A...531A..22P}, strongly suggesting that the nova erupted in an intermediate polar system and that residual nuclear surface burning is detected after the ejected envelope became optically thin for soft X-rays. The spin period measurement suggests an orbital period $\geq1.4$\,hrs for \ssd \citep[P$_{\mathrm{spin}} \sim 0.1$P$_{\mathrm{orb}}$;][]{2004ApJ...614..349N}.
However, the location of the source, either in the LMC or the Galaxy  is less clear.
Assuming the best-fit spectral parameters, the luminosity and size of the emission region at LMC distance are consistent with a white dwarf with nuclear burning on its surface near the Eddington limit. 
In this case \ssd could be similar to 1RXS\,J050526.3$-$684628 in the LMC, which exhibits 170\,s pulsations \citep{2020MNRAS.499.2007V}. The soft X-ray emission from both sources could be due to residual nuclear surface burning after a nova eruption. While the evolution of the X-ray luminosity of 1RXS\,J050526.3$-$684628 could be followed for about 30 years (first detections in \rosat data), \ssd was not detected during the \rosat all-sky survey, implying that at least the onset of the super-soft phase - and probably also the nova outburst - of \ssd occurred later than those of 1RXS\,J050526.3$-$684628. 

On the other hand, the spectral model for \ssd with one absorption component yields an absorption corrected 0.1--2.4\,keV luminosity of 5\ergs{33} for a distance of 1\,kpc.
The expected $L_{\rm x}$ of IPs range from 3\ergs{29} to 5\ergs{33} \citep[][]{2006ApJ...640L.167R} and with the above argument,  \ssd could be a soft IP located in the Galaxy at a distance $\leq1$\,kpc. The measured \nh is higher than the expected total Galactic column density in the direction of the source \citep[$\mathrm{N_{H}=N_{HI}+2N_{H_2}\approx2\times10^{21}~cm^{-2}}$,][]{2013MNRAS.431..394W}. 
This is indicative of obscuration local to the binary system which is common in IPs. Future deeper observations to firmly establish the presence/absence of the hard tail in the X-ray spectrum can help to ascertain/disregard the IP nature of this source.

\subsection{New Galactic magnetic CVs}

From our 4 new sources, which show super-soft X-ray emission,
at least 2 (\ssb and \ssc) are most likely located in our own Galaxy. 
\citet{2006A&A...452..431K} reported the \xmm detection of the SSS candidate RX\,J0059.4$-$7118 in the direction of the SMC 
and confirmed its super-soft X-ray spectrum, which is characterised by a blackbody component with a temperature of $\sim$90\,eV and an 
additional hard spectral component. 
Similarly, the high luminosity of RX\,J0059.4$-$7118 of $\sim$4\ergs{34}, if located at SMC distance, 
let the authors argue for a Galactic mCV. Given the similarities of RX\,J0059.4$-$7118 with 
\ssb and \ssc we support this conclusion and suggest the object to be a soft IP located in the Galaxy.
Another case with super-soft X-ray emission in the direction of the LMC is the double-degenerate candidate \xmmdd, 
which exhibits 1418\,s X-ray pulsations \citep{2017A&A...598A..69H}. 
For an LMC distance also this system exhibits a luminosity of around 5\ergs{34} and a radius for the emitting area of $\sim$180\,km.
The detection of a 23.6 minute periodic modulation in the optical and identification of the optical counterpart confirms a location of the binary system in the Galaxy \citep{2018A&A...617A..88R}. 

\ssb and \ssc, together with RX\,J0059.4$-$7118 and \xmmdd, make a significant contribution to the known population of such systems in our Galaxy \citep{2017A&A...598A..69H,2008A&A...489.1243A,2007ApJ...663.1277E}. 
The distance to \ssd is less clear, but it could be another IP system in the Galaxy.
\cite{2007ApJ...663.1277E} argue that whether IPs show a soft emission component mainly depends on the viewing geometry. Accretion curtains hide the soft emission from the surface of the white dwarf in the majority of IPs.
In addition to obscuration intrinsic to the binary system, the interstellar absorption along the line of sight plays an important role for the detection  of the soft emission.
A higher density of known sources in the direction of the Magellanic Clouds can likely be explained by the relatively low Galactic column density and the large number of existing X-ray observations sensitive to low energies.

\begin{acknowledgements}
We thank the referee for useful comments and suggestions which helped to improve the manuscript.
This work used observations obtained with \xmm, an ESA science mission with instruments and contributions directly funded by ESA Member States and NASA. The \xmm project is supported by the DLR and the Max Planck Society. 
This research has made use of the VizieR catalogue access tool, CDS,
Strasbourg, France. The original description of the VizieR service was
published in A\&AS 143, 23.
This work made use of data supplied by the UK Swift Science Data Centre at the University of Leicester.
The OGLE project has received funding from the National Science Centre, Poland, grant MAESTRO 2014/14/A/ST9/00121 to AU.
This work used data from \ero, the soft X-ray instrument on board \srg, a joint Russian-German science mission supported by the Russian Space Agency (Roskosmos), in the interests of the Russian Academy of Sciences represented by its Space Research Institute (IKI), and the Deutsches Zentrum f{\"u}r Luft- und Raumfahrt (DLR). The \srg spacecraft was built by Lavochkin Association (NPOL) and its subcontractors, and is operated by NPOL with support from the Max Planck Institute for Extraterrestrial Physics (MPE).
The development and construction of the \ero X-ray instrument was led by MPE, with contributions from the Dr. Karl Remeis Observatory Bamberg \& ECAP (FAU Erlangen-N{\"u}rnberg), the University of Hamburg Observatory, the Leibniz Institute for Astrophysics Potsdam (AIP), and the Institute for Astronomy and Astrophysics of the University of T{\"u}bingen, with the support of DLR and the Max Planck Society. The Argelander Institute for Astronomy of the University of Bonn and the Ludwig Maximilians Universit{\"a}t Munich also participated in the science preparation for \ero.
The \ero data shown here were processed using the \eSASS software system developed by the German \ero consortium. 

\end{acknowledgements}

\bibliographystyle{aa}
\bibliography{general}

\end{document}